\documentclass[
    ,final            
  ]
  {aipproc}

\layoutstyle{6x9}

\begin{document}

\title{Spectroscopy and pentaquarks at HERA}

\classification{13.60.Le, 13.60.Rj, 14.20.Lq, 14.40.Lb}

\keywords      {spectroscopy, pentaquarks}

\author{Leonid Gladilin\\
(on behalf of the H1 and ZEUS Collaborations)
}{
address={
DESY, ZEUS experiment, Notkestr. 85, 22607 Hamburg, Germany \\
E-mail: gladilin@mail.desy.de
}
,address={On leave from Moscow State University,
supported by the U.S.-Israel BSF} 
}

\begin{abstract}

Results of the H1 and ZEUS Collaborations on
spectroscopy
of light and charmed mesons and on
pentaquark searches,
obtained using the HERA I data,
are summarised.

\end{abstract}

\maketitle


\section{Introduction}

Light and charmed hadrons are produced copiously
in $ep$ collisions with a centre-of-mass energy of $318\,$GeV at HERA.
During first phase of HERA operation (1992-2000),
the H1 and ZEUS Collaborations accumulated data samples corresponding
to $\sim120\,$pb$^{-1}$ each.
The H1 and ZEUS
results on hadron spectroscopy and pentaquark searches are summarised
in this note.

\section{Spectroscopy of light and charmed mesons}

Inclusive photoproduction of $\eta$, $\rho^0$, $f_0(980)$ and $f_2(1270)$
mesons was measured at an average photon-proton centre-of-mass energy
$W=210\,$GeV~\cite{h1_f0f2}.
The differential cross sections for those mesons and for charged pions
as a function of $p_T+m$, where m is the meson's nominal mass, show
similar power-law behaviour. The results suggest a similar
mechanism of the mesons production in fragmentation processes.

Measurement of inclusive $K^0_s K^0_s$ production in deep inelastic
scattering (DIS) revealed a state at $1537\,$MeV,
consistent with $f^\prime_2(1525)$, and another
at $1726\,$MeV~\cite{zeus_f2f0}.
The state at $1726\,$MeV has a mass consistent with $f_0(1710)$,
and is found in a gluon-rich region of phase space.
This observation indicates that $f_0(1710)$ has
a sizeable gluonic component.

The production of excited charmed and charmed-strange mesons
was studied
using their decays to final states involving $D^{*\pm}$~\cite{zeus_dexc}.
The measured rates of $c$ quarks hadronising as $D^0_1$, $D^{*0}_2$ and
$D^\pm_{s1}$ mesons agree
with those obtained in $e^+e^-$ annihilations.
The measured value
of the helicity parameter for $D^\pm_{s1}$ mesons
is consistent with the observation of the
CLEO Collaboration that the spin-parity of the $D^\pm_{s1}$ is $1^+$.
A search for the radially excited $D^{*\prime\pm}$ meson revealed no signal.
The upper limit on
the product of the fraction of $c$ quarks hadronising as
a $D^{*\prime +}$ meson
and the branching ratio of the $D^{*\prime +}$ decay to
$D^{*+}\pi^+\pi^-$
was estimated to be $0.7\%$ ($95\%$ C.L.).

\section{Strange pentaquarks}

\begin{figure}
  \includegraphics[height=.32\textheight]{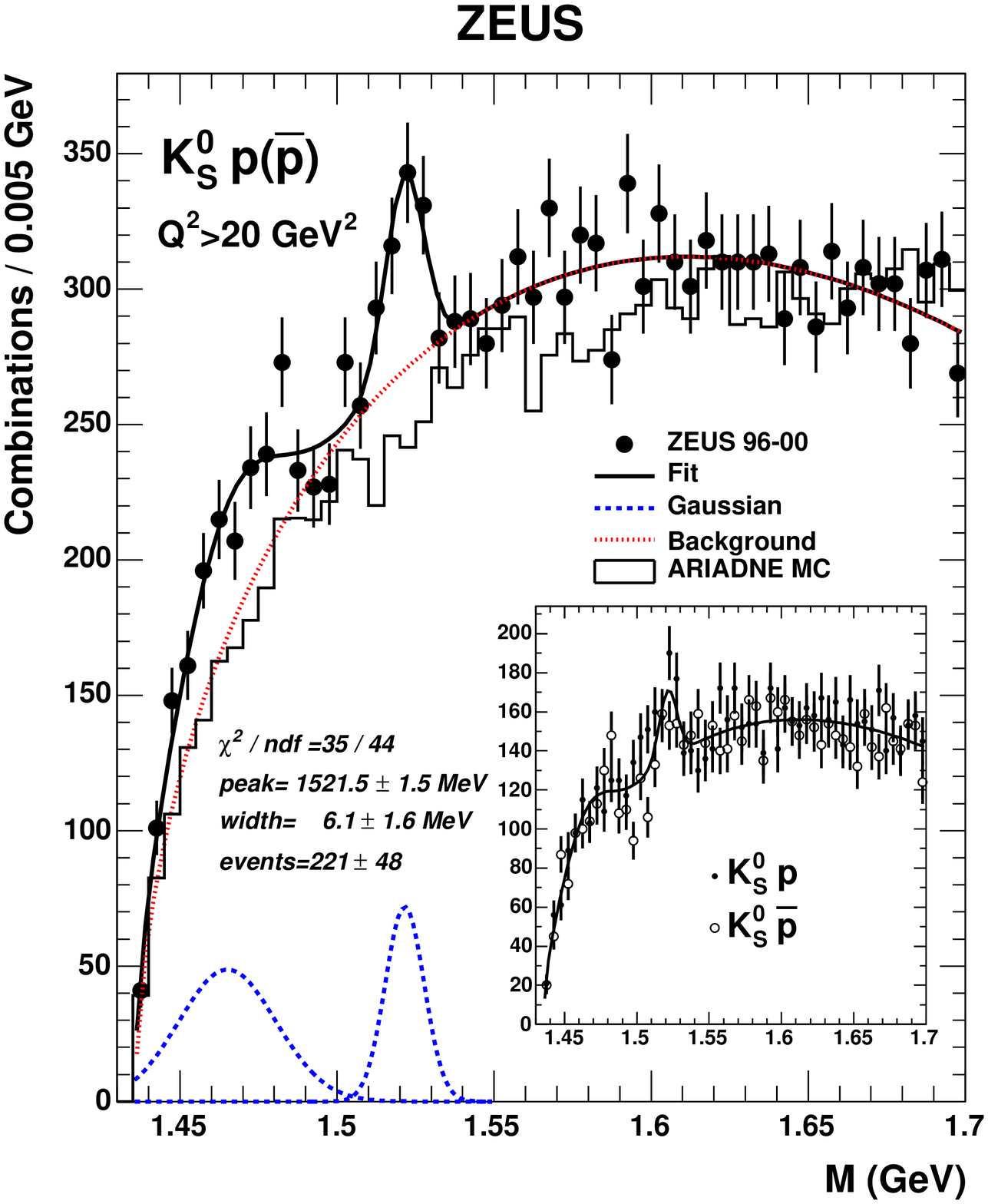}
\hspace*{-0.3cm}
  \includegraphics[height=.32\textheight]{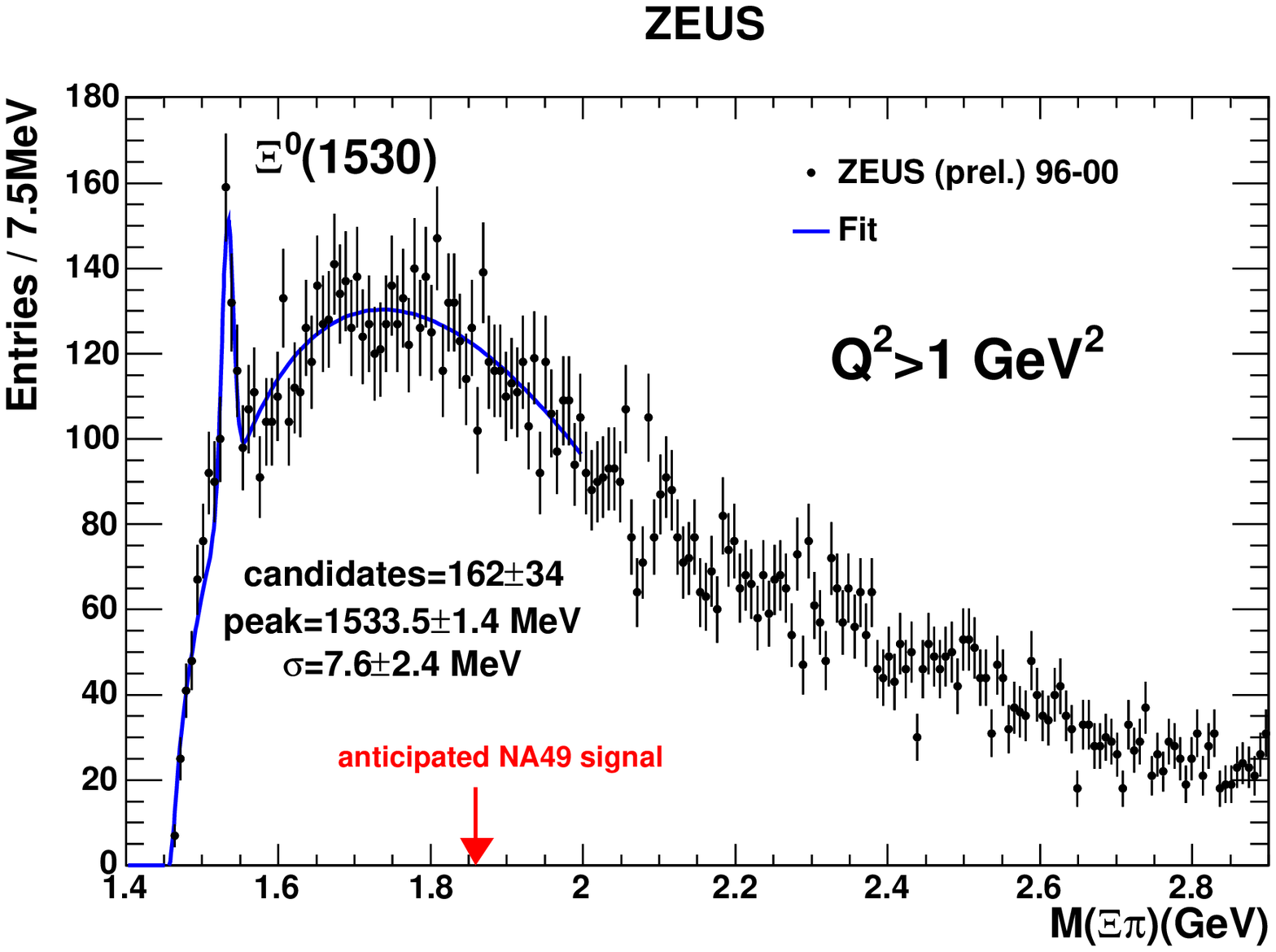}
  \caption{
Invariant mass spectrum for the $K^0_s p ({\bar{p}})$ combinations
for $Q^2>20\,$GeV$^2$ (left) and for $\Xi \pi$ combinations for
$Q^2>1\,$GeV$^2$ (right)
}
\end{figure}

A peak in the $K^0_s p ({\bar{p}})$ invariant mass spectrum around $1520\,$MeV
was observed in DIS by the ZEUS Collaboration~\cite{zeus_theta}.
In Fig.~1(left), the spectrum is shown for
exchanged photon virtuality
$Q^2>20\,$GeV$^2$.
The statistical significance of the signal varies between 3.9$\sigma$
and 4.6$\sigma$ depending upon the treatment of the background.
The signal is seen in both $K^0_s p$ and $K^0_s {\bar{p}}$ samples.
If the signal corresponds to the pentaquark
$\Theta^+$, this provides the first evidence for an anti-pentaquark
with a quark content ${\bar u}{\bar u}{\bar d}{\bar d}s$.
A ratio of the $\Theta^+$ and $\Lambda^0$ production cross sections 
was measured to be
$4.2\pm0.9({\rm stat.})^{+1.2}_{-0.9}({\rm syst.})\%$~\cite{zeus_thexs}.

The ZEUS Collaboration performed also a search
for two other pentaquarks, reported by the NA49 Collaboration, and
observed no signal in the $\Xi \pi$ invariant mass spectrum~\cite{zeus_xi}.
In Fig.~1(right), the spectrum is shown for $Q^2>1\,$GeV$^2$.
A clear peak with more than 160 $\Xi^0(1530)$ baryons indicates
that the statistical sensitivity of the search is similar
to that of the NA49 Collaboration.

\section{Charmed pentaquarks}

\begin{figure}
  \includegraphics[height=.46\textheight]{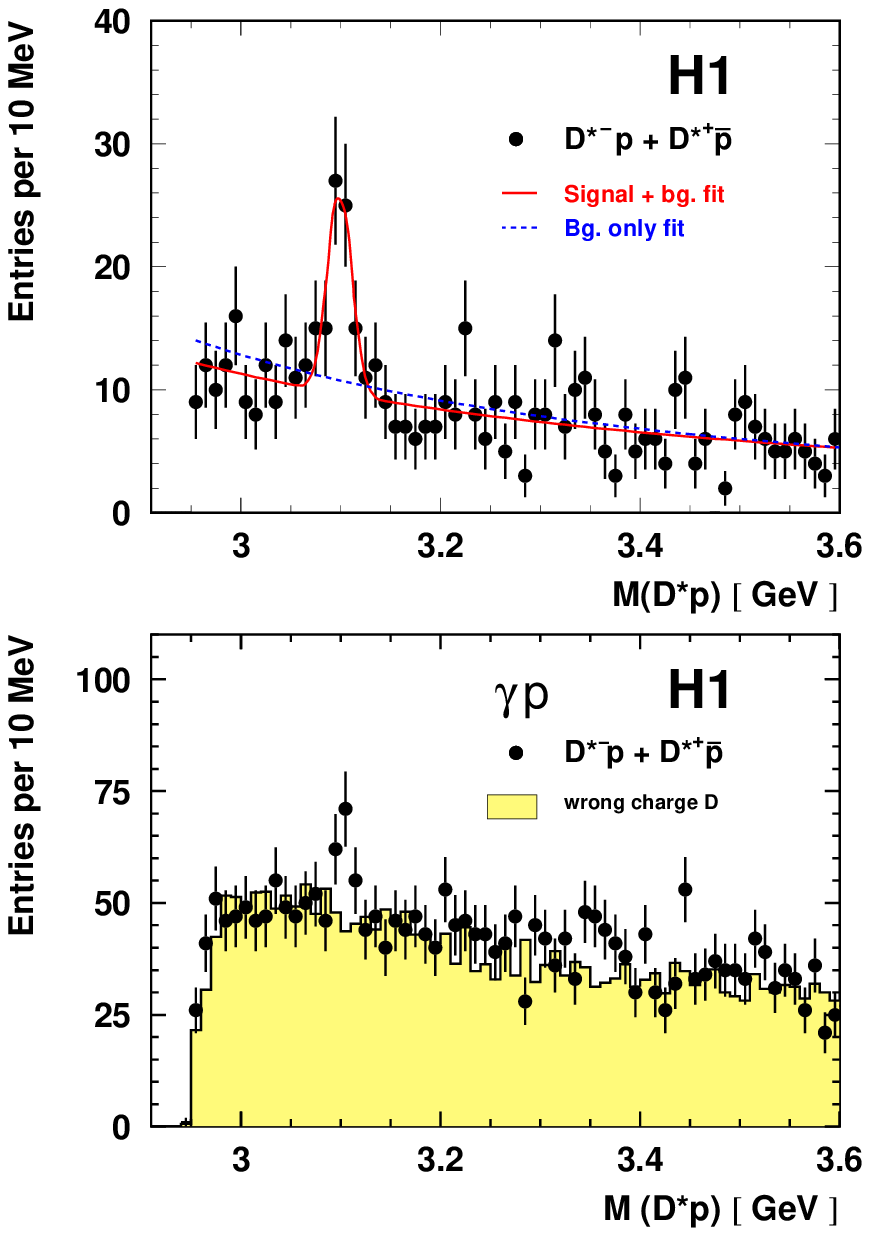}
\hspace*{-0.3cm}
  \includegraphics[height=.494\textheight]{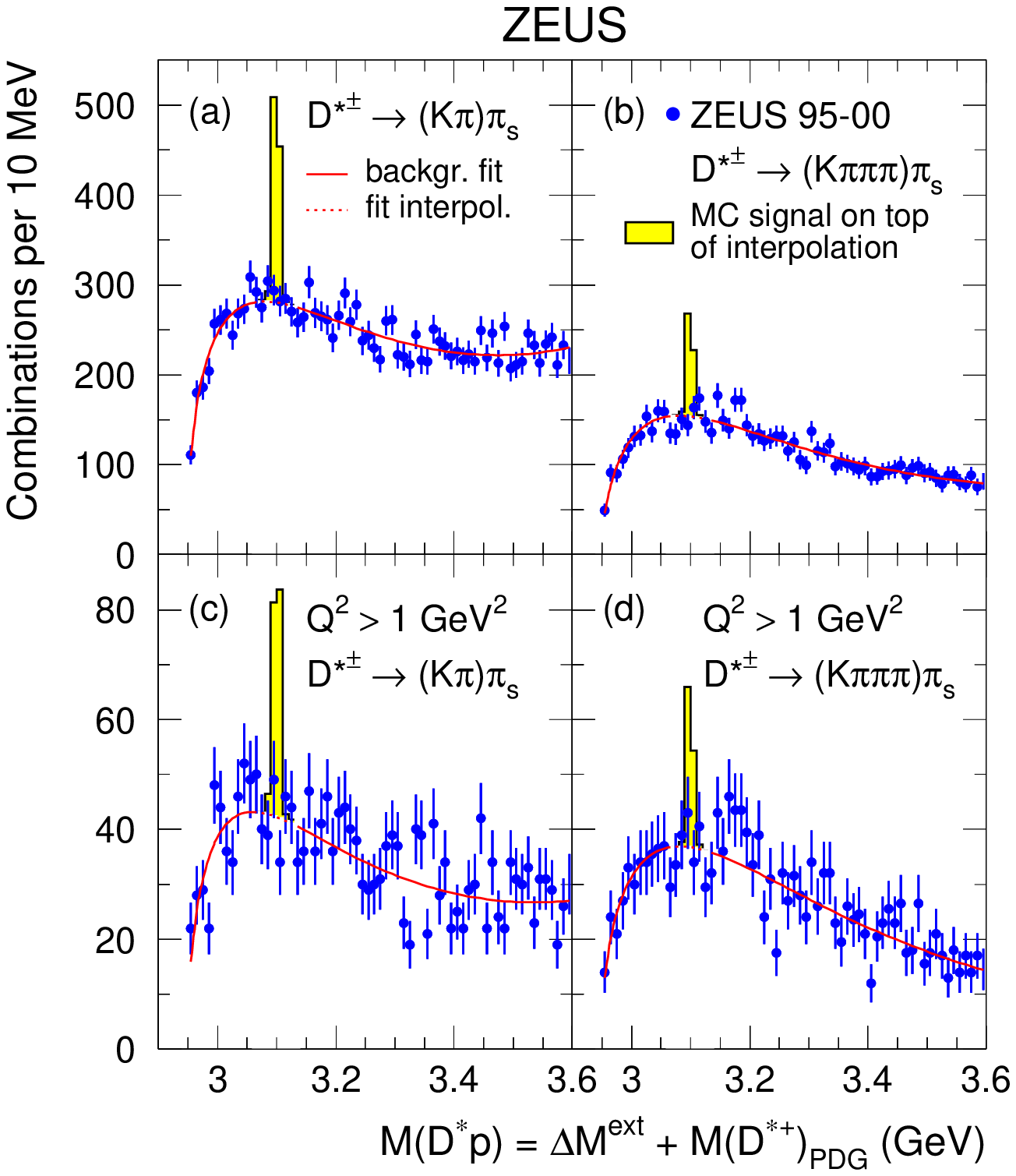}
  \caption{
The distributions of $M(D^{*\pm}p^\mp)$
obtained by the H1 Collaboration (left)
and by the ZEUS Collaborations (right)
}
\end{figure}

An observation of a candidate
for the charmed pentaquark state, $\Theta^0_c = uudd{\bar c}$,
decaying to $D^{*\pm}p^\mp$
was reported by the H1 Collaboration~\cite{h1_ch5q}.
Fig.~2(left) shows the $D^{*\pm}p^\mp$
invariant-mass distributions
in DIS with $Q^2>1\,$GeV$^2$ and in photoproduction
with smaller $Q^2$ values.
A fit of the signal in DIS yielded $50.6\pm11.2$ signal events and
the mass of $3099\pm3({\rm stat.})\pm5({\rm syst.})\,$MeV.
The observed resonance was reported to contribute
roughly $1\%$ of the $D^{*\pm}$ production rate
in the kinematic range studied in DIS.

The observation of the H1 Collaboration was challenged
by the ZEUS collaboration~\cite{zeus_ch5q}.
Using a larger sample of $D^{*\pm}$ mesons,
ZEUS observed no signature of the narrow resonance in
the $M(D^{*\pm}p^\mp)$ spectra shown in Fig.~2(right).
The Monte Carlo $\Theta^0_c$ signals normalised to $1\%$
of the number of reconstructed $D^{*\pm}$ mesons are shown
on top of the fitted backgrounds.
The upper limit on the fraction of $D^{*\pm}$ mesons
originating from $\Theta^0_c$ decays was evaluated to be 
$0.23\%$ ($95\%$ C.L.). The upper limit for DIS with $Q^2>1\,$GeV$^2$
is $0.35\%$ ($95\%$ C.L.).

\bibliographystyle{aipproc}   

\begin{thebibliography}{9}

\bibitem{h1_f0f2}
H1 Collab.,
Abstract 6-0184,
International Conference on High Energy Physics,
Beijing, China (ICHEP 2004), August 2004.
Available from http://www-h1.desy.de/.

\bibitem{zeus_f2f0}
ZEUS Collab., S.~Chekanov {\it et al.},
\emph{Phys. Lett.} \textbf{B 578} (2004) 33.

\bibitem{zeus_dexc}
ZEUS Collab.,
Abstract 854,
International Conference on High Energy Physics,
Osaka, Japan (ICHEP 2000), July-August 2000.\\
ZEUS Collab.,
Abstract 497,
International Europhysics Conference on High Energy Physics,
Budapest, Hungary (EPS 2001), July 2001.
Available from http://www-zeus.desy.de/.

\bibitem{zeus_theta}
ZEUS Collab., S.~Chekanov {\it et al.},
\emph{Phys. Lett.} \textbf{B 591} (2004) 7.

\bibitem{zeus_thexs}
ZEUS Collab,
Abstract 10-0273,
International Conference on High Energy Physics,
Beijing, China (ICHEP 2004), August 2004.
Available from http://www-zeus.desy.de/.

\bibitem{zeus_xi}
ZEUS Collab,
Abstract 10-0293,
International Conference on High Energy Physics,
Beijing, China (ICHEP 2004), August 2004.
Available from http://www-zeus.desy.de/.

\bibitem{h1_ch5q}
H1 Collab., C.~Atkas  {\it et al.},
\emph{Phys. Lett.} \textbf{B 591} (2004) 7.

\bibitem{zeus_ch5q}
ZEUS Collab., S.~Chekanov  {\it et al.},
DESY 04-164 (2004),
accepted by \emph{Eur. Phys. J.} \textbf{C}.

\end{thebibliography}

\end{document}